\UseRawInputEncoding

\documentclass[%
 reprint,
 superscriptaddress, % <--Group the authors
 amsmath,amssymb,
 aps,
 pra,
]{revtex4-2}

\usepackage{natbib}
\usepackage{placeins}

\usepackage{graphicx}% Include figure files
\usepackage{dcolumn}% Align table columns on decimal point
\usepackage{bm}% bold math
\usepackage{xcolor} % for \textcolor
\usepackage{xr}          % 若使用 hyperref，则用 xr-hyper
\begin{document}

\preprint{APS/123-QED}

\title{Breaking the bandwidth-efficiency trade-off in soliton microcombs via mode coupling}% Force line breaks with \\

\author{Yang Liu}
\thanks{These authors contributed equally.}
\affiliation{DTU Electro, Department of Electrical and Photonics Engineering, Technical University of Denmark, 2800 Kongens Lyngby, Denmark}

\author{Andreas Jacobsen}
\thanks{These authors contributed equally.}
\affiliation{DTU Electro, Department of Electrical and Photonics Engineering, Technical University of Denmark, 2800 Kongens Lyngby, Denmark}

\author{Thibault Wildi}
\affiliation{Deutsches Elektronen-Synchrotron DESY, Notkestr. 85, 22607 Hamburg, Germany}

\author{Yanjing Zhao}
\affiliation{DTU Electro, Department of Electrical and Photonics Engineering, Technical University of Denmark, 2800 Kongens Lyngby, Denmark}

\author{Chaochao Ye}
\affiliation{DTU Electro, Department of Electrical and Photonics Engineering, Technical University of Denmark, 2800 Kongens Lyngby, Denmark}

\author{Yi Zheng}
\affiliation{DTU Electro, Department of Electrical and Photonics Engineering, Technical University of Denmark, 2800 Kongens Lyngby, Denmark}

\author{Camiel Op de Beeck}
\affiliation{LIGENTEC SA, EPFL Innovation Park, 1024 Ecublens, Switzerland}

\author{Jos\'e Carreira}
\affiliation{LIGENTEC SA, EPFL Innovation Park, 1024 Ecublens, Switzerland}

\author{Michael Geiselmann}
\affiliation{LIGENTEC SA, EPFL Innovation Park, 1024 Ecublens, Switzerland}

\author{Kresten Yvind}
\affiliation{DTU Electro, Department of Electrical and Photonics Engineering, Technical University of Denmark, 2800 Kongens Lyngby, Denmark}

\author{Tobias Herr}
\affiliation{Deutsches Elektronen-Synchrotron DESY, Notkestr. 85, 22607 Hamburg, Germany}

\author{Minhao Pu}
\thanks{mipu@dtu.dk}
\affiliation{DTU Electro, Department of Electrical and Photonics Engineering, Technical University of Denmark, 2800 Kongens Lyngby, Denmark}

\date{\today}% It is always \today, today,
             %  but any date may be explicitly specified

\begin{abstract}
Dissipative Kerr solitons in optical microresonators have emerged as a powerful tool for compact and coherent frequency comb generation. Advances in nanofabrication have allowed precise dispersion engineering, unlocking octave-spanning soliton combs  that are essential for applications such as optical atomic clocks, frequency synthesis, precision spectroscopy, and astronomical spectrometer calibration. 
However, a key challenge hindering their practical deployment is the intrinsic bandwidth-efficiency trade-off: achieving broadband soliton generation requires large pump detuning, which suppresses power coupling and limits pump-to-comb conversion efficiencies to only a few percent. Recent efforts using pulsed pumping or coupled-resonator architectures have improved efficiency to several tens of percent, yet their bandwidths remain below one-tenth of an octave, inadequate for applications demanding wide spectral coverage.
Here, we overcome this limitation by harnessing mode interactions between spatial modes within a single microresonator. The mode hybridization creates an additional power-transfer channel that supports large pump detuning while maintaining strong pump-to-resonator coupling, enabling broadband soliton formation at substantially reduced pump power. Using this approach, we demonstrate an octave-spanning soliton microcomb with a record pump-to-comb conversion efficiency exceeding 50\%. These results resolve the fundamental bandwidth-efficiency dilemma in soliton microcombs and paves the way toward fully-integrated, high-efficiency, ultrabroad comb sources for next-generation photonic systems.
\end{abstract}

%\keywords{Suggested keywords}%Use showkeys class option if keyword
                              %display desired

\maketitle
\let\thefootnote\relax
\footnotetext{* These authors contributed equally}
\footnotetext{\dag\ tobias.herr@desy.de}
\footnotetext{\ddag\ mipu@dtu.dk}

%\tableofcontents

\section{Introduction}

Dissipative Kerr soliton (DKS) microcombs \cite{Herr2014TemporalMicroresonators,Kippenberg2018} provide compact, coherent and highly scalable optical frequency combs for applications in metrology \cite{trocha2018ultrafast}, spectroscopy \cite{Suh2016, yu2018silicon} and telecommunications \cite{Marin-Palomo2017}. For practical deployment, two performance metrics are especially critical. The pump-to-comb conversion efficiency governs power consumption and system scalability, and the optical bandwidth enables capabilities such as self-referencing for optical clocks \cite{Wu2025} and broadband molecular spectroscopy \cite{yu2018silicon}. 

A fundamental challenge arises from the intrinsic bandwidth-efficiency trade-off. In the standard soliton model, the soliton bandwidth scales with pump-resonator detuning \cite{Yi2015}. Achieving large detuning requires substantial pump power, but large detuning simultaneously reduces the effective power coupled into the resonator, restricting pump-to-comb conversion efficiencies to only a few percent \cite{Bao2014,Jang2021,Zhang2024,Yang2024}. This trade-off fundamentally limits the achievable performance of current DKS microcombs.

A wide range of strategies has been explored to improve the efficiency or bandwidth of soliton microcombs. To enhance pump-to-comb conversion efficiency, prior efforts include tailoring the cavity-to-bus coupling to improve soliton out-coupling \cite{bao2014nonlinear,jacobsen2025high}, employing pulsed pumping to enhance the intracavity pump-soliton temporal overlap\cite{Li2022} and developing device-level strategies such as pump recycling in coupled resonators \cite{xue2019super}, interferometric back-coupling \cite{Boggio2022}, photonic crystal reflectors \cite{Zang2025}, and mode-shifting in photonic molecules \cite{Helgason2023b}. These approaches have achieved efficiencies of several tens of percent, yet their optical bandwidth remain limited. In parallel, dispersion engineering has been used to phase-match dispersive waves and enlarge the attainable comb span, enabling octave-spanning and spectrally flattened soliton spectra under favorable conditions \cite{Brasch2016d,pfeiffer2017octave,Moille2023a}. However, such dispersion-engineered bandwidth enhancements remain fundamentally constrained by the underlying bandwidth–efficiency trade-off dynamics. Additional broadening mechanisms, including spectral broadening \cite{del2016phase,lamb2018optical} or spectral translation \cite{zhang2020spectral}, can further extend the spectral span beyond the native soliton bandwidth, but at the cost of increased system complexity and power consumption. Until now, no existing approach simultaneously achieves high efficiency and ultra-broad bandwidth in a single soliton microcomb. Overcoming this limitation is crucial for realizing practical, power-efficient, and frequency-flexible integrated comb systems.

\begin{figure*}[htpb]
	\begin{center}
		\includegraphics[width=1\linewidth]{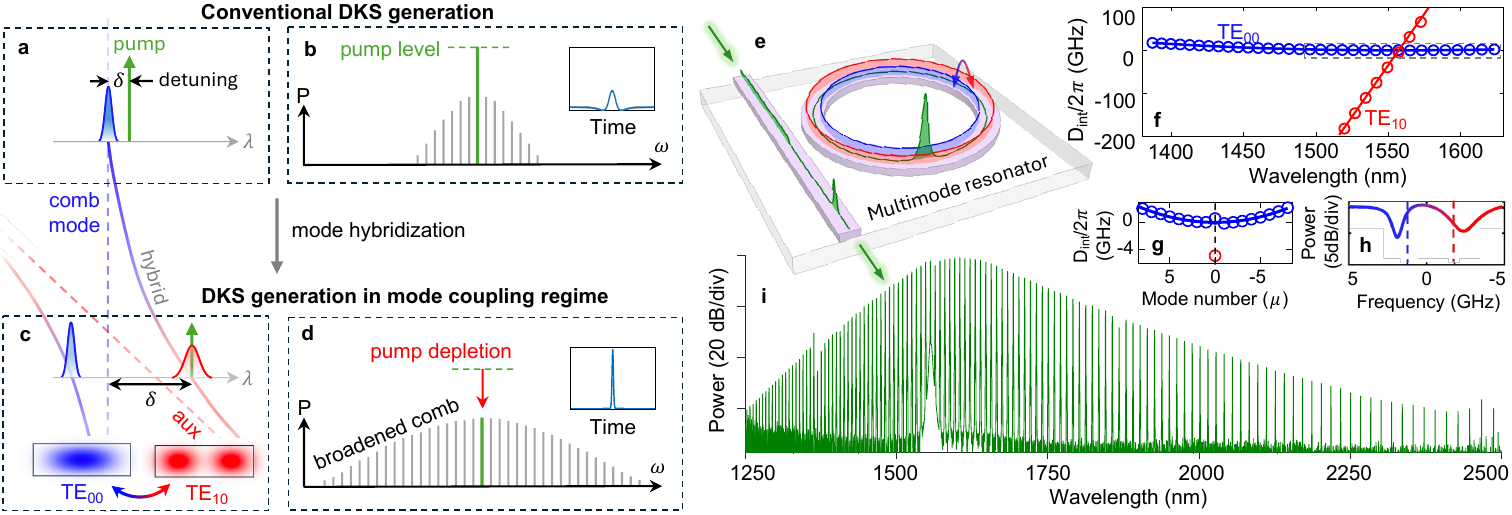}
		\caption{\textbf{Broadband and efficient dissipative Kerr soliton generation enabled by mode coupling.} (a) Conventional dissipative Kerr soliton (DKS) generation in a microresonator, where the pump red-detuning slightly from the comb mode. (b) Typical DKS comb spectrum showing substantial residue pump power exceeding the comb peak. Inset: temporal soliton profile over a round-trip time. (c) DKS generation in the mode coupling regime, where an auxiliary mode (TE$_{10}$) hybridizes with the comb mode (TE$_{00}$). Pumping the hybridized resonance larger effective detuning and enhances power transfer via mode hybridization. (d) DKS comb spectrum under mode coupling. Extended detuning broadens the spectrum, while resonant pumping at the hybridized mode leads to efficient pump depletion. Inset: soliton pulse trace with higher peak power and reduced background. (e) Schematic of a multimode microresonator supporting coupling between TE$_{00}$ and TE$_{10}$ modes. (f) Measured integrated dispersion (D$_{int}$) of both modes with respect to the free spectra range (FSR) of the TE$_{00}$ mode at 1556 nm, showing an avoided mode crossing (AMX). (g) Zoomed-in of the dispersion near the AMX region (dashed box in (f)). (h) Measured transmission spectrum near the AMX, where mode coupling shifts both resonances from their unperturbed positions (marked by the vertical blue and red lines for TE$_{00}$ and TE$_{10}$, respectively). (i) Measured DKS comb spectrum at 50 mW pump power showing an octave-spanning bandwidth and nearly complete pump depletion.}
		\label{fig:Fig1}
	\end{center}
\end{figure*}

In this work, we address this challenge by introducing strong mode coupling between distinct spatial modes within a single microresonator. The resulting hybridized mode forms an energy-transfer pathway that continuously channels pump power into the soliton-forming mode, even under the large pump detuning required for broadband soliton generation. Critically, pumping the hybridized mode provides access to new soliton operation regime that cannot be reached with conventional pumping schemes, while simultaneously enhancing pump depletion and thereby boosting pump-to-comb conversion efficiency. This unified mechanism stabilizes broadband soliton formation at reduced pump power and over detuning ranges far exceeding those tolerated in conventional single-mode operation. Using this approach, we demonstrate a single-soliton, octave-spanning microcomb with a record pump-to-comb conversion efficiency of 53\%, achieving both ultra-broad bandwidth and high efficiency within a single device.

\section{Results}

\textbf{Operation principle} Fig. \ref{fig:Fig1}(a-d) schematically compare the conventional pumping scheme (referred to as single-mode pumping scheme hereafter) and coupled-mode pumping scheme in strong coupling regime. For a fixed pump power P, both the soliton power and the soliton bandwidth scale with $\sqrt\delta$, where $\delta$ is the detuning between pump and the comb mode \cite{Herr2014TemporalMicroresonators,Yi2015}. In the single-mode pumping scheme shown in Fig. \ref{fig:Fig1}(a), the maximal achievable detuning $\delta_\mathrm{max}$ of the soliton state is fundamentally limited by the cavity resonance, because the intracavity pump power rapidly diminishes as the pump is tuned away from resonance, eventually preventing soliton formation. Therefore, increasing the pump power $P$ increases the maximal achievable detuning $\delta_\mathrm{max}$ of a soliton state. Finally, a tradeoff arises, because the pump-to-comb conversion efficiency scales with $1/\sqrt{P}$, while soliton spectrum bandwidth scales with $\sqrt{P}$. This trade-off fundamentally hinders the simultaneous maximization of both efficiency and bandwidth in the single-mode pumping scheme. Fig. \ref{fig:Fig1}(b) schematically shows a typical soliton spectrum and the pulse of a soliton state in the single-mode pumping scheme, where only modest bandwidth and efficiency can be achieved. Moreover, power of the pump is much higher than that of the comb lines in the spectrum as the pump is far detuned from resonance, so practical applications typically require output-side filtering to suppress the pump.

To address this limitation, we propose the coupled-mode pumping scheme, in which an auxiliary mode is introduced to strongly coupled with the comb mode, as shown in Fig. \ref{fig:Fig1}(c). When the separation of the two intrinsic modes (dashed lines) is small, the actual modes (solid lines) would repel each other, causing a significant deviation from the intrinsic modes, a phenomenon called avoided mode crossing (AMX). Fig. \ref{fig:Fig1}(c) schematically shows the configuration of the modes in a coupled-mode pumping scheme, where the auxiliary mode (red mode) is located on the long-wavelength side of the comb mode (blue mode). The strong mode coupling leads to mode hybridization between the two modes, opening a channel to transfer the power of the resonant auxiliary mode to the comb mode. Therefore, the actual comb mode cavity power is dependent on both coupling from the external bus waveguide and the auxiliary mode. Finally, maximal detuning of a soliton state is no longer only dependent on the pump power, but also on spatial mode coupling strength and the separation of the two modes. By operating the soliton with pumping a well separated strongly-coupled auxiliary mode, the maximal detuning can be significantly increased. Fig. \ref{fig:Fig1}(d) schematically shows a soliton spectrum and the pulse of a soliton state in the coupled-mode pumping scheme. The significantly improved comb detuning allows for simultaneous improvement of both efficiency and bandwidth, breaking the conventional trade-off between efficiency and bandwidth. Moreover, locating the pump at the center of the auxiliary mode together with the coupling condition shifted by a powerful soliton state allows for a strong pump depletion of the pump power in the output spectrum.

\textbf{Device and soliton performance.} We consider a $\mathrm{Si_{3}N_{4}}$ multimode microresonator fabricated by Ligentec supporting both $\mathrm{TE_{00}}$ and $\mathrm{TE_{10}}$ modes for the coupled-mode pumping scheme, as shown in Fig.~\ref{fig:Fig1}(e) (see Supplementary Section I). The mode profiles of the $\mathrm{TE_{00}}$ and $\mathrm{TE_{10}}$ modes, overlaid with a waveguide of dimensions $0.8\times1.54\ \mu\mathrm{m}^{2}$, are shown in Fig.~\ref{fig:Fig1}(c). The multimode microresonator is coupled to a single-mode bus waveguide through a point coupler. Therefore, both $\mathrm{TE_{00}}$ and $\mathrm{TE_{00}}$ modes in the cavity exchange energy with the $\mathrm{TE_{00}}$ mode in the bus waveguide through evanescent field coupling \cite{pfeiffer2017coupling}. The $\mathrm{TE_{00}}$ and $\mathrm{TE_{10}}$ mode have average intrinsic quality factors of $1.8\times10^{6}$ and $7.0\times10^{5}$, respectively, and average loaded quality factors of $4.1\times10^{5}$ and $7.3\times10^{4}$, respectively. The $\mathrm{TE_{00}}$ mode acts as the mode for comb generation, and the $\mathrm{TE_{00}}$ mode is the auxiliary mode. The free spectral ranges (FSRs) of the $\mathrm{TE_{00}}$ and $\mathrm{TE_{10}}$ modes are 985 GHz and 951 GHz, respectively. Fig.~\ref{fig:Fig1}(f) shows the measured integrated dispersion of the $\mathrm{TE_{00}}$ and $\mathrm{TE_{10}}$ at the pump resonance at around 1556.15 nm. The $\mathrm{TE_{00}}$ mode exhibits anomalous dispersion with $D_2/2\pi=62.4\ \mathrm{MHz}$, while the $\mathrm{TE_{10}}$ mode has normal dispersion with $D_2/2\pi=-89.0\ \mathrm{MHz}$. Fig.~\ref{fig:Fig1}(g) shows the zoom-in of the integrated dispersion at around the pumped resonance, where the strong mode coupling causes a significant deviation of the actual $\mathrm{TE_{00}}$ mode from the dispersion fitting curve. Fig.~\ref{fig:Fig1}(h) shows the measured transmission spectrum of the pumped resonance, with the blue and red lines indicating the resonance frequency of the intrinsic $\mathrm{TE_{00}}$ and $\mathrm{TE_{10}}$ modes, respectively. From the measured integrated dispersion and transmission spectrum, we can extract the coupling rate between the $\mathrm{TE_{00}}$ and $\mathrm{TE_{10}}$ modes to be $K_c/2\pi=|K_c|/2\pi\cdot\exp(i\varphi)=3.2\exp{(i11\pi/18)}$ GHz (See Supplementary Section II).

Fig.~\ref{fig:Fig1}(i) shows the soliton spectrum generated by pumping the hybridized auxiliary $\mathrm{TE_{10}}$ mode with approximately 50-mW of pump power. The hybridized auxiliary mode enables large comb detuning, yielding an octave-spanning soliton comb with an unprecedented $>$50\% pump-to-comb conversion efficiency. Simultaneously, the soliton is generated when pumping at the center of the $\mathrm{TE_{10}}$ mode, allowing the $\mathrm{TE_{10}}$ mode to act as an effective filter that nearly depletes the pump and eliminates residual pump light at the output spectrum.

\begin{figure*}[htpb]
	\begin{center}
		\includegraphics[width=0.8\linewidth]{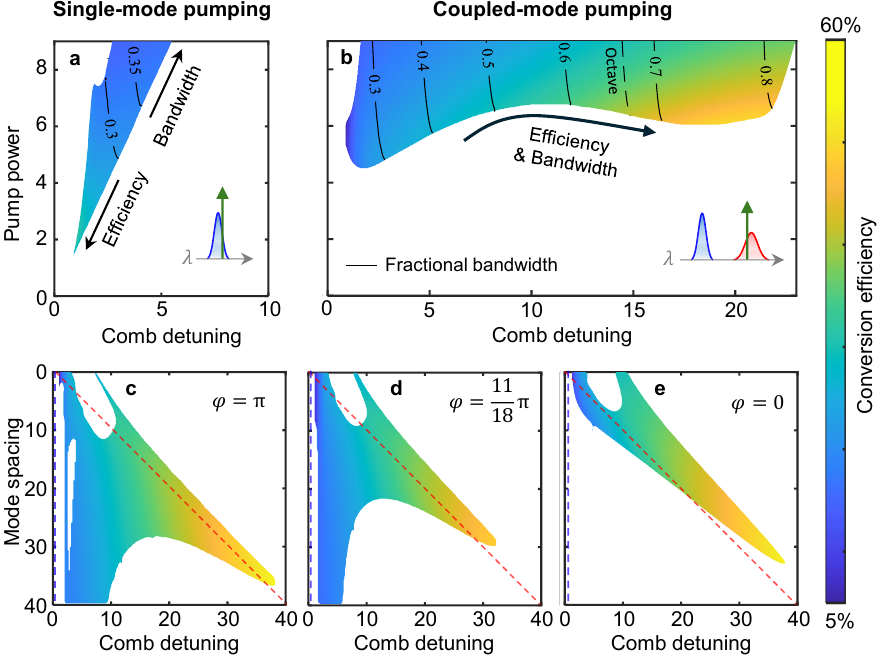}
		\caption{\textbf{Mode coupling extends soliton existence range and enhances conversion efficiency.} Single-soliton existence and pump-to-comb conversion efficiency maps for microcombs under (a) single-mode pumping and (b) coupled-mode pumping schemes. The normalized comb detuning is defined with respect to the unperturbed resonance of the comb mode (i.e., in the absence of coupling). Color shading represents conversion efficiency, and solid contour lines indicate the 70-dB fractional comb bandwidth. Arrows highlight directions toward improving either efficiency or bandwidth. Single-mode pumping reveals a trade-off between bandwidth and efficiency, while coupled-mode pumping introduces a localized regime where both can be simultaneously optimized. (c-d) Single-soliton existence and efficiency map under coupled-mode pumping at constant pump power of 8 for different phase of coupling: (c) $\varphi=\pi$, (d) $\varphi=\frac{11}{18}\pi$ (the measured phase), and (e) $\varphi=0$. $\varphi$ is defined in coupling strength $K_c = |K_c|\exp(i\varphi)$. The blue and red dashed lines trace the intrinsic resonance of comb mode and auxiliary mode, with the auxiliary mode kept red-detuned from the comb mode.}
		\label{fig:Fig2}
	\end{center}
\end{figure*}

\begin{figure*}[htpb]
	\begin{center}
		\includegraphics[width=1\linewidth]{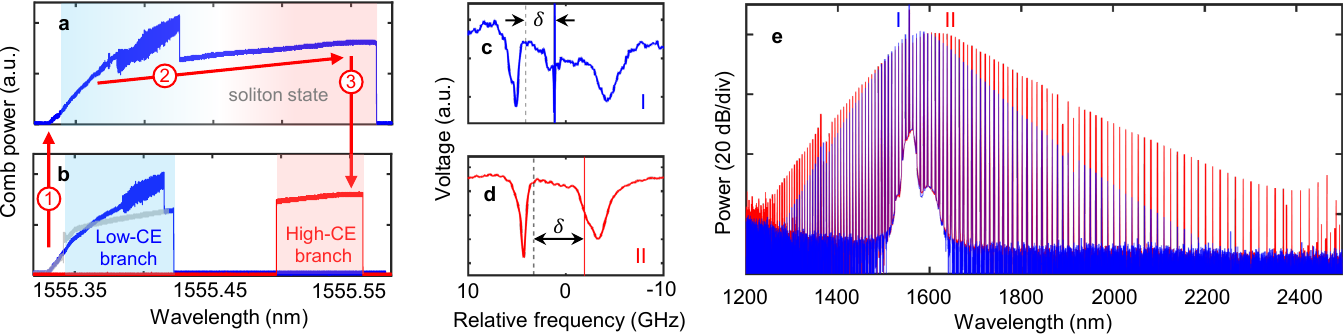}
		\caption{\textbf{Accessing single-soliton state with large effective comb detuning at low pump power.} (a, b) Measured comb power traces as the pump wavelength is tuned at high (a) and low (b) pump power under the coupled-mode pumping scheme. Blue traces show the comb power evolution during forward tuning pump across the coupled-mode resonance. Color-shaded regions indicate the existence ranges of single-soliton states. At low pump power, two distinct soliton existence regions emerge, corresponding to low-efficiency and high-efficiency branches. Forward tuning at low power (b) accesses only the low-efficiency branch. The high-efficiency branch can be reached by: 1) increasing the pump power beyond the threshold where the two branches merge; 2) forward tuning into the single soliton state at large effective comb detuning; and 3) subsequently reducing the pump power while maintaining the soliton state. The red trace shows the comb power evolution when tuning the pump after reaching the high-efficiency branch from a high-power state. (c, d) Measured pump detuning for soliton combs operating in the low-efficiency (c) and high-efficiency (d) branches identified in (b). Dashed lines mark the position of the unperturbed TE$_{00}$ resonance. (e) Measured soliton comb spectra at low pump power. Blue and red spectra correspond to the detuning cases shown in (c) and (d), respectively. The extended effective comb detuning ($\delta$) in the high-CE branch enables ultra-broadband comb generation.}
		\label{fig:Fig3}
	\end{center}
\end{figure*}

\textbf{Numerical investigation of soliton performance in coupled-mode pumping scheme.} To gain a theoretical understanding of the coupled mode pumping scheme, we carried out numerical calculations by solving a coupled LLE (See Methods). We first compare single-mode and coupled-mode pumping schemes by numerically investigating the conversion efficiency and bandwidth of a single soliton state across the parameter space of pump power and detuning. We begin by analyzing the single-mode pumping scheme in Fig.~\ref{fig:Fig2}(a), using the measured Q factors of the $\mathrm{TE_{00}}$ mode in the simulation. We can see that the the max detuning of the single soliton state scales linearly with pump power as expected, matching well with previously reported results \cite{godey2014stability}. From the colormap, we see that the conversion efficiency is limited to around 25.6\% when the soliton is operated with the least pump power, while the 70-dB fractional bandwidth is also limited to 0.4 at the highest simulated pump power. This trade-off limits the overall performance of the soliton state with both efficiency and bandwidth constrained. 

% Fig.~\ref{fig:Fig2}(d) shows the simulated single-soliton spectrum at the black circle in the parameter space of single-mode pumping scheme, corresponding to the highest conversion efficiency, while 70-dB bandwidth is around 50 THz with fractional bandwidth of around 0.15. 

We then map the existence map for the coupled-mode pumping scheme shown in Fig.~\ref{fig:Fig2}(b), In this case, the simulation uses the experimentally extracted Q factors of the $\mathrm{TE_{00}}$ mode (comb mode) and $\mathrm{TE_{10}}$ mode (auxiliary mode), and also the coupling strength between them. The auxiliary mode is located on the red-detuned side of the comb mode with unperturbed mode spacing to be 20 times the comb mode linewidth. We can see that single-soliton existence range is dramatically broadened across nearly all pump power levels compared to Fig. \ref{fig:Fig2}(a). Moreover, there are two soliton existence area branches appearing; one lies close to comb mode with minimum power of around \textcolor{red}{4.5} (referred to as low-efficiency branch), and the other one lies close to the auxiliary mode with threshold of around \textcolor{red}{6} (referred to as high-efficiency branch). Most importantly, the region of maximum conversion efficiency shifts from the low-power, small-detuning low-efficiency branch to the high-efficiency branch of higher power and larger detuning as the arrow in Fig. \ref{fig:Fig2}(b) indicates.

We also numerically investigate how the spectral spacing between the two modes affects the maximum achievable detuning for a given pump power. Fig.~\ref{fig:Fig2}(d) plots the single-DKS existence map versus detuning and unperturbed mode spacing for a fixed normalized pump power of 8 while the other parameters are the same as linear characterization. Two soliton branches still appear: one lies close to the intrinsic comb mode (blue dashed line) and gradually converges toward the soliton existence region of the single-mode pumping scheme as the mode spacing increases, while the other is far red-detuned and follows the intrinsic mode of the auxiliary mode (red dashed line). When the two modes are relatively near each other (mode spacing $<21$ in Fig.~\ref{fig:Fig2}(d)), these branches merge. For larger separations (mode spacing $>21$ in Fig.~\ref{fig:Fig2}(d)) a gap opens between them, because the optical power within the gap is insufficient to sustain solitons. We can see that the high-efficiency soliton branch persists until the mode spacing reaches approximately 28.5, with its maximum detuning being about six times that of the low-efficiency soliton branch.

We next investigate how the phase of the mode-coupling coefficient, $\varphi$, affects the soliton diagram in the parameter space of comb detuning and mode spacing. Figures~\ref{fig:Fig2}(c--e) show three representative cases, corresponding to $\varphi=\pi$, $\varphi=\tfrac{11}{18}\pi$ (the measured value), and $\varphi=0$. First, we find that the high-efficiency branches persist for all coupling phases and continuously follow the intrinsic resonance of the auxiliary mode. Their maximal achievable detuning and conversion efficiency remain nearly unchanged across different $\varphi$. This indicates that the existence and performance of the high-efficiency soliton branch are largely insensitive to the coupling phase. In contrast, the low-efficiency branch exhibits a strong dependence on $\varphi$: it gradually shrinks as $\varphi$ decreases from $\pi$ toward $0$ and eventually disappears. This behavior originates from interference between the two excitation channels of the comb mode---the direct coupling from the bus waveguide and the intra-cavity mode coupling. The coupling phase $\varphi$ sets the interference condition. When $\varphi=0$, the two channels interfere destructively when the pump is near the comb-forming mode, substantially reducing the intra-cavity power in that mode. Although the high-efficiency branch exists for all phases, the ability to \emph{access} this branch during a pump sweep does depend on $\varphi$. Accessing the soliton state requires first initiating parametric oscillation as the pump approaches the comb mode from the short-wavelength side. When destructive interference suppresses the intra-cavity power (as in Fig.~\ref{fig:Fig2}(e) with $\varphi=0$), parametric oscillation cannot be triggered, making it difficult to reach the high-efficiency soliton branch through standard tuning methods. In contrast, for phases where the low- and high-efficiency branches merge (Fig.~\ref{fig:Fig2}(c,d)), the intra-cavity power is sufficient to initiate parametric oscillation, enabling straightforward access to the soliton state. In our multimode microresonator platform, the naturally realized coupling phase already lies in a favorable regime for initiating parametric oscillation during pump sweeping. Moreover, engineered designs, such as angular gratings \cite{ulanov2024synthetic}, may be employed to further tailor $\varphi$ if needed.

\begin{figure*}[htpb]
	\begin{center}
		\includegraphics[width=1\linewidth]{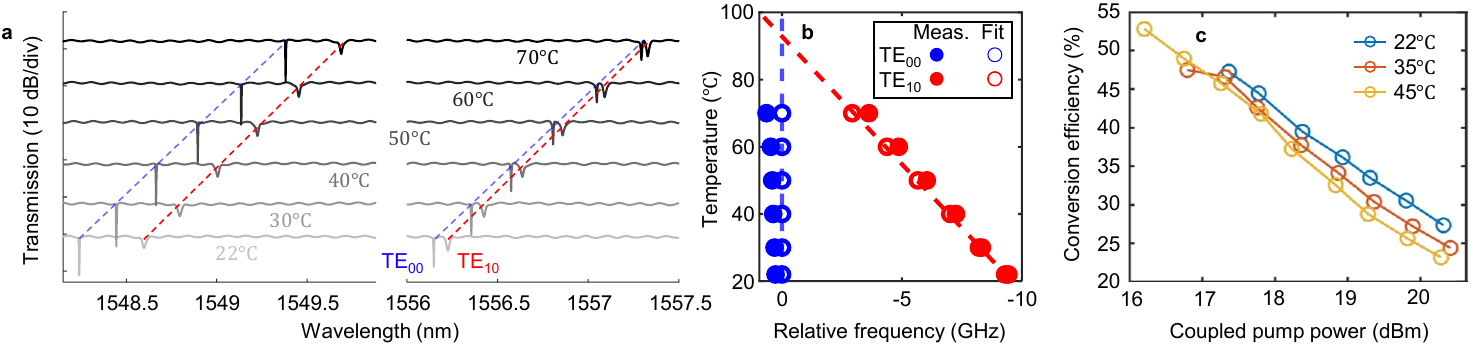}
		\caption{\textbf{Enhancing pump-to-comb conversion efficiency via tailored mode coupling.} (a) Transmission spectrum of the microresonator at different temperatures. (b) Measured integrated dispersion (D$_{int}$) of the TE$_{00}$ mode at different temperatures, demonstrating that the mode coupling strength (evident as the coupling-induced resonance shift of the comb mode) can be significantly tuned. (c) Measured pump-to-comb conversion efficiency (CE) as a function of pump power for different initial mode coupling strengths. CE is evaluated at the maximum achievable effective comb detuning under each pumping condition. A maximum CE of approximately 53\% is obtained by increasing the mode coupling through temperature tuning to $45\,^\circ\mathrm{C}$.}
		\label{fig:Fig4}
	\end{center}
\end{figure*}

\textbf{Access to the high-efficiency soliton branch.} We pumped the microresonator with a tunable laser amplified by an erbium-doped fiber amplifier. The residual pump was suppressed using a fiber Bragg grating, and the filtered comb light was detected on a photodiode to monitor the comb power, while the full optical spectrum was simultaneously recorded with an optical spectrum analyzer (see Supplementary Section~III). We first forward tuned the pump from the blue side at a relatively low on-chip power of 57~mW. As shown by the blue trace in Fig.~\ref{fig:Fig3}(b), only a narrow single-soliton step is observed during forward tuning. Performing a backward tuning from this step reveals the full soliton existence region near the comb-forming mode (grey curve), which corresponds to the low-efficiency branch in Fig.~\ref{fig:Fig2}(b).

As the pump power is increased, the soliton existence range during forward tuning broadens dramatically, as shown in Fig.~\ref{fig:Fig3}(a). This broadening arises from the merging of the low-efficiency and high-efficiency soliton branches, in agreement with our numerical simulations. At low pump power, these two branches remain separated, which prevents direct access to the high-efficiency branch via a simple forward tuning process. However, by first entering a large-detuning operating point at higher pump power and subsequently reducing the pump power while maintaining the detuning, we are able to reach the high-efficiency branch even though it is not directly accessible at 57~mW, as illustrated by the red trace in Fig.~\ref{fig:Fig3}(a). The resulting single-soliton dynamics are markedly different from those observed when pumping a resonance with negligible mode coupling in the same device, where only a narrow multi-soliton step appears and a single soliton can be generated only by performing an additional backward tuning process~\cite{Guo2017}.

The soliton spectra associated with operation on the low-efficiency (blue) and high-efficiency (red) branches are shown in Fig.~\ref{fig:Fig3}(e). The high-efficiency branch exhibits a substantially broadened spectrum extending over an octave. Fig.~\ref{fig:Fig3}(c-d) show the measured detuning for the two branches. In the high-efficiency case, the pump is located on the auxiliary mode and achieves a significantly larger effective comb detuning from the comb-forming mode than in the low-efficiency case. Remarkably, under these conditions we obtain a pump-to-comb conversion efficiency of approximately 50\%.

Beyond enabling a larger accessible detuning, operating the pump on the auxiliary mode provides additional advantages associated with the fundamental limit of pump-to-comb efficiency. Although the efficiency is influenced by multiple parameters, its ultimate ceiling is set by the degree of pump depletion. When the pump is injected into the comb-forming mode, soliton operation typically requires the pump to be far red-detuned, which substantially suppresses the intracavity pump buildup and constrains the pump depletion to a relatively low level. In contrast, pumping the auxiliary mode allows the pump to operate much closer to its resonance, enabling much stronger intracavity buildup and making it possible for the system to approach a significantly higher depletion level than would be inaccessible with conventional pumping of the comb mode.

A second benefit arises from the coupling condition of the auxiliary mode. In our microresonator, the $\mathrm{TE_{10}}$ mode is over-coupled in the linear regime. Once soliton formation begins, the strong nonlinear energy transfer from the pump to the comb lines manifests as an effective nonlinear loss acting on the pump. This additional loss increases the total dissipation experienced by the pump and naturally drives the coupling condition from over-coupling toward critical coupling~\cite{xue2017microresonator}. Operating near critical coupling maximizes the fraction of pump power that is both coupled into the resonator and available for nonlinear conversion, allowing the system to further increase the efficiency limit. This behavior provides a practical design guideline: the auxiliary mode should be engineered to be over-coupled in the linear regime so that nonlinear conversion during soliton operation can pull it toward critical coupling. As a consequence, even though the auxiliary $\mathrm{TE_{10}}$ mode exhibits a linear extinction ratio of less than 5~dB, the pump is nearly fully depleted in the soliton state, as shown in Fig.~\ref{fig:Fig1}(i).

\textbf{Conversion efficiency improvement.} Numerical simulations indicate that the spacing between the modes fundamentally limits the maximum achievable comb detuning, and thus the conversion efficiency, as shown in Fig.~\ref{fig:Fig2}(c-e). To experimentally probe this effect, we thermally tune the device and measure the resulting changes in the modal structure. Figure~\ref{fig:Fig4}(a) shows the transmission spectra at different temperatures. The $\mathrm{TE_{00}}$ and $\mathrm{TE_{10}}$ modes exhibit different thermal shift rates of 2.92~GHz/$^\circ\mathrm{C}$ and 2.79~GHz/$^\circ\mathrm{C}$, respectively. As shown in Fig.~\ref{fig:Fig4}(b), the relative frequency separation between the modes decreases at higher temperature, accompanied by larger deviations from their intrinsic resonances, indicating stronger modal hybridization.

We pump the device at different temperatures to investigate how this thermally induced change of mode spacing affects conversion efficiency. Figure~\ref{fig:Fig4}(c) summarizes the conversion efficiency near the largest accessible detuning on the high-efficiency branch for various temperatures and pump powers. A clear trend is observed: at a fixed pump power, lower temperatures yield higher conversion efficiencies. This results from the larger uncoupled-mode spacing at low temperature, which expands the soliton existence range and enables larger comb detunings, thereby increasing the attainable comb power.

At elevated temperatures, however, the situation differs. The reduced mode spacing leads to a smaller detuning range, and the generated comb power is correspondingly lower. Nonetheless, solitons can be sustained with significantly lower pump power due to the strengthened modal hybridization, which enhances energy transfer from the auxiliary mode to the comb mode. As a result, although the absolute comb power decreases at high temperature, the required pump power decreases even more. This disproportionate reduction in pump power leads to an increase in the conversion efficiency, which is defined as the ratio between comb power and pump power.

These observations reveal two different mode-spacing-dependent effects: larger mode spacing at low temperature increases the maximum achievable detuning and thereby improves the absolute comb power, while stronger modal hybridization at high temperature reduces the soliton sustaining power more rapidly than it reduces the comb power, thereby improving the conversion efficiency ratio. This interplay highlights mode-spacing control as a promising strategy for optimizing both conversion efficiency and comb power in soliton microcombs.

\section{Discussion}

In this work, we have demonstrated that a coupled-mode pumping scheme enables soliton microcombs that simultaneously achieve high efficiency and broad bandwidth. Intermodal spatial coupling provides a controlled pathway that directs pump power from an auxiliary mode into the soliton-forming mode under large detuning, supporting stable soliton operation through simple single-pump tuning. This mechanism allows octave-spanning soliton generation while maintaining strong pump-to-comb conversion, enabling efficient pump depletion and comb output powers exceeding tens of milliwatts. The resulting combination of wide bandwidth, high efficiency, and substantial output power establishes a robust platform for applications such as millimeter-wave synthesis \cite{tetsumoto2021optically} and optical atomic clocks \cite{Wu2025}.

Optical bandwidth and energy efficiency are two critical performance metrics of soliton microcombs that must be improved for many applications. Here, we demonstrate that a coupled-mode pumping scheme can simultaneously enhance both. We identify intermodal spatial coupling as a mechanism that provides power from an auxiliary mode for soliton sustaining at large comb detuning. High-performance soliton generation is thus achieved through straightforward single-pump tuning and power control. We successfully realize an octave-spanning soliton microcomb with a conversion efficiency of approximately 53\%. Moreover, the high-efficiency soliton operates under high pump power, enabling the generation of a powerful comb output exceeding tens of milliwatts. Such high efficiency, large output power, and broad bandwidth promise a high-performance soliton microcomb for applications including millimeter-wave oscillators \cite{tetsumoto2021optically} and optical atomic clocks \cite{Wu2025}. The coupled-mode pumping concept is broadly applicable and offers a general route to improving soliton performance across diverse microcomb platforms. Additional enhancements are readily accessible. Engineering the intermodal coupling rate through tailored resonator designs \cite{ulanov2024synthetic} may further expand soliton bandwidth and improve conversion efficiency. Likewise, integrating this strategy with complementary device-level optimizations such as improved bus-to-cavity coupling or the use of ultra-high-Q ($>10^7$) microresonators could unlock even higher levels of performance.\\

\textbf{Acknowledgments}\\
This work is supported by European Research Council (REFOCUS 853522, ELASTIC 101213805), Swiss State Secretariat for Education, Research and Innovation (SERI), European Innovation Council (CSOC 101047289), Danish National Research Foundation (SPOC ref. DNRF123), Independent Resarch Fund Denmark (ifGREEN 3164-00307A), and Innovationsfonden (GreenCOM 2079-00040B, EDOCS 4354-00020B).\\
%\\
%\textbf{Author contributions}\\
%Y. L. and M. P. conceived the idea. C. Y. and Y. Z. designed the device. C. O. B., J. C. and M. G. fabricated the sample. A. J., Y. L. and Y. J. Z. performed the experiment.  Y. L., T. W. and T. H. contributed to the theoretical understanding. A. J., Y. L., Y. J. Z., T. W., T. H. and M. P. analyzed the data. K. Y., T. H. and M. P. supervised the project. Y. L., A. J., T. H. and M. P. wrote the manuscript. All authors reviewed the results and provided feedback on the manuscript.\\
%\\
%\textbf{Competing interests}\\
%Y.L., K.Y., and M.P. are listed as inventors on a patent application that broadly covers aspects related to this work. The remaining authors declare no competing interests.\\
%\\

% The \nocite command causes all entries in a bibliography to be printed out
% whether or not they are actually referenced in the text. This is appropriate
% for the sample file to show the different styles of references, but authors
% most likely will not want to use it.
% \nocite{*}
\bibliographystyle{unsrt}
\bibliography{References/references_backup.bib}% Produces the bibliography via BibTeX.

\subsection{\label{sec:level2}LLE modeling with coupled mode equations }

We simulate the nonlinear process in the multimode microresonator through two linearly coupled Lugiato–Lefever equations (LLEs) \cite{fujii2018analysis}:
\small
\begin{multline*}
\frac{dA}{dt}
+ i\sum_{n=2}^{\infty}\frac{(-i)^nD_{n}}{n!}\frac{\partial^nA}{\partial \phi^n}
- ig|A|^2A -ir2g|B|^2A = \\-\Bigl(\frac{K_1}{2} + i\omega_1 \Bigl) A
+ i\frac{K_c}{2}B - \frac{\sqrt{K_\mathrm{ex1}K_\mathrm{ex2}}}{2}B+\sqrt{\frac{K_{\text{ex1}} P_\mathrm{in}}{\hbar \omega_0}}
\end{multline*}

\begin{multline*}
\frac{dB}{dt}
 -ig|B|^2B - ir2g\sum_\mu |A_\mu|^2B = -\Bigl(\frac{K_2}{2} + i\omega_2 \Bigl) B \\
 + i\frac{K_c^*}{2}A_{\mu}\delta_{\mu0} - \frac{\sqrt{K_\mathrm{ex1}K_\mathrm{ex2}}}{2}A_{\mu}\delta_{\mu0}
+ \sqrt{\frac{K_{\text{ex2}} P_\mathrm{in}}{\hbar \omega_0}}
\end{multline*}

\normalsize

where $A(\phi,t)$ is the slowly varying field amplitude of the comb mode in the microresonator, $\phi$ is the co-rotating angular coordinate in the microresonator, $B$ is the mode amplitude of the auxiliary mode near the pumped comb mode, $A_{\mu}$ is the mode amplitude of the $\mu$-th comb mode, $D_n$ is the coefficients of microresonator integrated dispersion, $K_1=K_{\mathrm{i1}}+K_{\mathrm{ex1}}$ is the total decay rate of the comb mode, including the intrinsic decay rate $K_{\mathrm{i1}}$ and external coupling decay rate $K_{\mathrm{ex1}}$, $K_2=K_{\mathrm{i2}}+K_{\mathrm{ex2}}$ is the total decay rate of the auxiliary mode, including the intrinsic decay rate $K_{\mathrm{i2}}$ and external coupling decay rate $K_{\mathrm{ex2}}$, $g$ is the nonlinear coupling coefficient, $P_\mathrm{in}$ is the pump power, $\omega_1$ and $\omega_2$ are the pump detuning relative to the intrinsic resonant frequency of the comb mode and auxiliary mode, respectively, and $\delta_{\mu0}$ is the Kronecker delta, $r$ represents the relative ratio of the intermodal to the intramodal nonlinear interaction strength. The linear coupling between comb mode and auxiliary is divided into two parts: the $K_c$ is the linear coupling rate between the comb mode and the auxiliary mode happening within the microresonator, while the term with $- \sqrt{K_\mathrm{ex1}K_\mathrm{ex2}}/{2}$ represents the coupling happening at the cavity-to-bus coupler region \cite{nazemosadat2021switching}. AMXs far from the pump are less of interest in this scheme, and the focus in this work is on pumping an AMX. Therefore, we introduce only one auxiliary mode that couples with one comb mode near the pump. The auxiliary mode is assumed to follow a homogeneous solution, which is valid when no comb is generated in the auxiliary spatial mode. For simplicity, we take the same nonlinear coupling rate for the comb mode and the auxiliary mode.

The comb mode used in this simulation is set with $K_{\mathrm{i1}}/2\pi = 104\ \mathrm{MHz}$ and $K_{\mathrm{ex1}}/2\pi = 173\ \mathrm{MHz}$. The auxiliary mode is set with $K_{\mathrm{i2}}/2\pi = 471\ \mathrm{MHz}$ and $K_{\mathrm{ex2}}/2\pi = 2177\ \mathrm{MHz}$. Fig. \ref{fig:Fig2}(a) and Fig. \ref{fig:Fig2}(c) are simulated with only comb mode. Fig. \ref{fig:Fig2}(b), Fig. \ref{fig:Fig2}(d) and Fig. \ref{fig:Fig2}(e) are simulated with intra-cavity mode coupling between the two modes to be $K_{\mathrm{c}}/K_1 = 11.7\exp{(i\frac{110}{180}\pi)}$. The other parameters are $g =17.55 \mathrm{\ s^{-1}}$, $D_2/2\pi = 62.44\ \mathrm{MHz}$, $\mathrm{FSR} = 1000\ \mathrm{GHz}$, $r=1$ for simplification of calculation. 

The single-DKS existence region is investigated using two approaches: time-domain parameter sweeping and numerical continuation \cite{godey2014stability}.

The time-domain adiabatic parameter-sweeping method include the following step: (1) A grid with a resolution of 0.1 in the normalized power and normalized detuning is generated in the parameter space;
(2) An analytical expression of the single soliton state is used to seed the model, and at least one grid point where the single soliton state exists is identified;
(3) The identified grid point(s) are used as starting points in the parameter space. The neighboring grid points are then examined by propagating the model from the starting point(s) to the neighboring points within 5 photon lifetime;
(4) After reaching a neighboring point, the simulation is continued for an additional 44.5 photon lifetimes to allow the intracavity state to stabilize;
(5) A coherence check within the final 5 photon lifetime is performed to determine the intracavity state. If a single DKS state is found, the corresponding grid point is added as a new starting point in the next iteration, and the numerical solution of the corresponding DKS state is used as the new seed.
(6) Steps (3)-(5) are repeated until no new single soliton states are discovered.

The numerical continuation method include the following step: (1) A grid with a resolution of 0.2 in the normalized uncoupled mode spacing and normalized detuning is generated in the parameter space;
(2) An analytical expression of the single-soliton state is used to seed the model, and at least one grid point supporting a single-soliton solution is identified;
(3) The identified grid point(s) are used as starting points in the parameter space to solve for the stability solution in neighboring grid points;
(4) For each computed steady-state solution, the Jacobian matrix of the linearized Lugiato–Lefever equation is evaluated, and its eigenvalues are used to determine whether the solution is dynamically stable;
(5) Steps (3)-(4) are repeated iteratively until no additional stable single-soliton states are found in previously unexplored grid points.

\end{document}